\documentclass[aps,pre,showpacs,groupedaddress,
               showkeys,amsmath,amssymb,twocolumn]{revtex4}

\usepackage{graphicx}
\usepackage{bm}
\usepackage{bbm}

\usepackage{epic}
\usepackage{eepic}
\usepackage{pifont}

\usepackage{nicefrac}
\input epsf
\input rotate
\usepackage{graphicx}
\begin{document}
\draft

\title{Dynamics and rheology of a dilute suspension of vesicles:
higher order theory}
\author{Gerrit Danker, Thierry Biben$^{*}$, Thomas Podgorski, Claude Verdier and Chaouqi Misbah}
\affiliation{Laboratoire de Spectrom\'etrie Physique, UMR, 140
avenue de la physique, Universit\'e Joseph Fourier, and CNRS,  38402
Saint Martin d'Heres, France\\
$^*$ Present address: Universit\'e de Lyon, Laboratoire PMCN,
Universit\'e Claude Bernard-Lyon I
     et CNRS, 43 bvd du 11 novembre 1918, 69622 Villeurbanne, France}

\email[]{chaouqi.misbah@ujf-grenoble.fr}
\date{\today}
\begin{abstract}
Vesicles under shear flow exhibit various dynamics: tank-treading
($tt$), tumbling ($tb$) and vacillating-breathing ($vb$). A
consistent higher order theory reveals  a direct bifurcation from
$tt$ to $tb$ if $C_a\equiv \tau \dot\gamma $ is small enough
($\tau$= vesicle relaxation time towards equilibrium shape,
$\dot\gamma$=shear rate). At larger $C_a$ the $tb$  is preceded by
the $vb$ mode. For $C_a\gg 1$ we recover the leading order original
calculation, where the $vb$ mode coexists with $tb$. The consistent
calculation reveals several quantitative discrepancies with recent
works, and points to new features. We  analyse rheology and find
that the effective viscosity exhibits  a minimum at $tt-tb$ and
$tt-vb$ bifurcation points.

\end{abstract}


\maketitle

Vesicles are closed membranes suspended in an aqueous medium. They
constitute an interesting starting model for the study of dynamics of
real cells, such as red blood cells. The study of their rheology should
capture some essential features of blood rheology.

Under a linear shear flow, a vesicle (where the membrane is in its
fluid state) is known to exhibit a tank-treading ($tt$) motion,
while its long axis makes an angle, $\psi<\pi/4$, with the flow
direction\cite{Kraus96,Seifert99}. In the presence of a viscosity
contrast $\lambda=\eta_{in}/\eta_{o}$ ($ \eta_{in}$ and $\eta_{o}$
are the internal and external viscosities, respectively), $\psi$
decreases until it vanishes at a critical value of
$\lambda=\lambda_c$. For a small enough $C_a\equiv \tau \dot\gamma $
($\tau$ is the relaxation time towards the equilibrium shape in the
absence of an imposed flow, $\dot\gamma$=shear rate) the $tt$
exhibits a saddle-node bifurcation towards tumbling
($tb$)\cite{Biben03}.

Recently, a new type of motion has been predicted\cite{Misbah06},
namely a vacillating breathing ($vb$) mode: the vesicle's long axis
undergoes an oscillation (or vacillation) around the flow direction,
while the shape executes a breathing motion. Shortly after this
theoretical prediction, an experimental report on this type of mode
has been presented\cite{Kantsler06} (trembling denomination was used
there) and in \cite{Mader06} a qualitatively similar motion called
"transition motion" in the vicinity of the $tt-tb$ transition has
been observed. Nevertheless, a detailed experimental study of this
$vb$ mode would be interesting but has not been reported yet. Since
then, works providing further understanding\cite{Petia}, or
 attempting\cite{Lebedev,Gompper} to extend the original theory\cite{Misbah06}
%(denomination adopted there are respectively trembling and swinging)
to higher order deformation (with the aim  to account for the
experimental observation\cite{Kantsler06}) have been presented.
Interesting features have emerged\cite{Gompper,Lebedev} regarding
the behavior of the $vb$ mode as a function of $C_a$ ($C_a$ scales
out in the leading theory\cite{Misbah06}).

The first aim of this communication is to present the result of the
consistent theory regarding the higher order calculation. We find
significant differences with  a recent work\cite{Lebedev,Gompper}
regarding the form of the evolution equation. This implies, in
particular, that the location of the boundaries separating the
various three regimes in parameters space is significantly affected.
%The original theory could be viewed
%formally as corresponding\cite{Misbah06} to the limit of a very
%large $C_\kappa$.
%We shall use the abbreviations $tt$
%(tank-treading), $tb$ (tumbling), and $VB$.
%
%In this paper, we present a consistent theory taking into account
%higher order deformations. Unlike Refs. \cite{Lebedev,Gompper}, we
%develop a consistent analysis taking into account the hydrodynamic
%response to the same order as the membrane bending forces.
%We
%present a phase diagram of the bifurcation structure of the $tt$,
%$vb$ and $tb$ regimes. We extend  the rheological study presented

A second important report is to investigate how the effective
viscosity derived recently in\cite{Danker07} is affected by the
higher order deformation.
% We
%report on the effective viscosity of a dilute suspension.
For  a small enough $C_a$ the effective viscosity  of the suspension
(as a function of $\lambda$) still exhibits a cusp singularity at
the $tt-tb$ bifurcation\cite{Danker07}, while the cusp becomes a
smooth minimum when $C_a$ is high enough, namely when  the $tt-vb$
bifurcation occurs.
%At low enough $C_\kappa$, the cusp seems to be
%washed out by the inclusion of higher order terms of the vesicle
%shape.

The vesicle suspension is submitted to a linear shear flow ${\bf
V_0}=(\dot\gamma y, 0,0)$. The fluid inside and outside the vesicle
is described by the Stokes equations, with the following boundary
conditions: (i) continuity of the (normal and tangential) velocity
across the membrane, (ii) continuity of normal and tangential stress
at the membrane, (iii) membrane incompressibility. The basic
technical spirit can be found
elsewhere\cite{Frankel70,Cox69,Seifert99,Misbah06,Petia}. Here, we
merely focus on the results. Lengths are reduced by the vesicle
radius $r_0$ ($r_0$ designates the radius of a sphere having the
same volume). The shape of the vesicle can be written in the general
case as an infinite series on the basis of spherical harmonics
${\cal Y}_{nm}$
\begin{equation}
r=1+\epsilon\sum_{n=0}^\infty \sum_{m=-n}^n F_{nm}(t){\cal
Y}_{nm}(\theta,\phi) \label{shape}
\end{equation}
where $\epsilon$ is a small parameter expressing a small deviation
from a sphere, $\theta$ and $\phi$ are the usual angles in spherical
coordinates, and $F_{nm}(t)$ is a time dependent amplitude (to be
determined) of the corresponding spherical harmonic. We shall set
$\epsilon=\sqrt{\Delta}$, where $\Delta$
 is the membrane excess area defined by $A=4\pi
+\Delta$, $A$ being the dimensionless area of the vesicle.

Using the expression of spherical harmonics in terms of Cartesian
coordinates, $r_i$, we have
\begin{equation}
\sum_{m=-2}^2 F_{2m}(t)\,{\cal Y}_{2m}(\theta,\phi)
=\sum_{i,k=x,y,z} 3 f_{ik}(t)\, r_ir_k\label{shape2},
\end{equation}
$f_{ik}$ are linear combinations of $F_{2m}$.
% The algebra has proven
% be  more convenient with $f_{ik}$.
Since a shear flow induces a shape deformation from a sphere which
involves only second order harmonics (i.e. $n=2$), only ${\cal
Y}_{2m}$ enters the calculation\cite{Seifert99,Misbah06}.

In recent works\cite{Gompper,Lebedev} interesting attempts to deal
with the higher order deformation (the next step beyond the leading
order theory\cite{Misbah06}) have been made. However, these authors
take into account only  the higher order contribution in the
Helfrich bending force. We find that there are other contributions
(especially stemming from the corresponding hydrodynamics response)
which are at least of the same order. This is the first goal of this
communication. The next goal is to analyze the far-reaching
consequences of the present theory. Finally we examine some
rheological implications.

The method follows the same strategy as in\cite{Misbah06}, and the
technical details will be reported elsewhere, while  we focus here
on the main results. The extension of the leading order
calculation\cite{Misbah06} to higher order yields, for the
amplitudes of the shape functions $f_{ij}$, the following evolution
equation (please compare with Eq.(3) in Ref.\cite{Danker07})
% ind that the evolution
%equation for the amplitudes
%$f_{ij}$ takes the following form:
\begin{eqnarray}
  \label{eq_ffij}
  \epsilon\,\frac{{\cal D}f_{ij}}{{\cal D}t} &=&
  \frac{20\,e_{ij}}{23\lambda+32} -
  \frac{24\,(Z_0+6\bar\kappa)}{23\lambda+32}\,f_{ij} \\
  &&\hspace{-8ex}{}+\epsilon\bigg[\frac{4800}{7}\frac{\lambda-2}{(23\lambda+32)^2}\,{\cal
  S}d[f_{ip}e_{pj}] \nonumber\\
  &&\hspace{-8ex}{}+\frac{288}{7}\frac{(49\lambda+136)\,Z_0 +
  (432\lambda+1008)\,\bar\kappa}{(23\lambda+32)^2}\,{\cal
  S}d[f_{ip}f_{pj}]\bigg]\nonumber,
\end{eqnarray}
where $e_{ij}=[\partial _i v_j+ \partial_j v_i]/2$ and ${\cal
S}d[b_{ij}] = \frac{1}{2}[b_{ij} + b_{ji} -
\frac{2}{3}\delta_{ij}b_{ll}]$.  $Z_0(t)$ is the isotropic component
of the membrane tension and is determined by the condition of
constant surface excess area. This constraint provides a relation
between the $f_{ij}'s$ which expresses the fact that the shape
evolution equations comply with the available excess area. Making
use of (\ref{eq_ffij}) fixes then $Z_0$ in terms of  $f_{ij}$ (and
other parameters, like $\Delta$, $C_a$). This relation is lengthy
and will be listed in an extended paper.
$\bar\kappa=\kappa\sqrt{\Delta}$, where $\kappa$ is the membrane
rigidity. Note that even though $\sqrt{\Delta }$ has served formally
to make an expansion of the full equations, it turns out that the
amplitudes are small enough so that even for $\Delta \simeq 1$ the
perturbative scheme makes a sense. Our understanding of this fact is
that $\Delta=1$ corresponds to only $8\%$ of  relative excess area.

The quantity ${\cal D/ D}t$ entering Eq.~(\ref{eq_ffij}) is the Jaumann
(or corotational) derivative defined as
\begin{equation}
{{\mathcal D} \mathbf{M}\over {\mathcal D} t}= {D\mathbf{M}\over
Dt}+{1\over 2} [ {\mathbf \omega}{\mathbf M} -{\mathbf M}{\mathbf
\omega}]
 \label{jaumann}
\end{equation}
where ${\bf M}$ is any second order tensor,  $D/Dt$ is the usual
material derivative, and ${\bf \omega }=({\bf \nabla v} - {\bf\nabla
v}^T)/2$ is the vorticity tensor. A fully consistent calculation
induces highly nonlinear terms.

Let us first discuss the evolution  equation (\ref{eq_ffij}) and
compare it to recent studies\cite{Misbah06,Gompper,Lebedev}. For
that purpose, it is convenient to use another set of variables,
namely the orientation angle $\psi$ of the vesicle in the plane of
the shear and the shape amplitude $R$ by setting
$F_{22}=Re^{-2i\psi}$ \cite{Misbah06}. We may alternatively, instead
of $R$, use the definition $R/ 2\sqrt{\Delta}=\cos(\Theta )$, as in
\cite{Lebedev}. We expand the full equation in powers of $f_{ij}$
and retain terms up to the higher (fifth) order in a consistent
manner. We then perform a straightforward conversion of variables in
terms of $\psi$ and $\Theta$. We find for $\Theta$ and $\psi$ the
following equations:
\begin{eqnarray}
  \label{eq_alpha}
&&T\partial_t\Theta =- S\sin\Theta\sin 2\Phi + \cos 3\Theta +
\epsilon\,\Lambda_1\,S\sin 2\Phi\ \times \nonumber
\\
 \times && (\cos 4\Theta + \cos 2\Theta) +
   \epsilon\,\Lambda_2\,S\sin 2\Phi\cos 2\Theta +....
 \\
  T\partial_t\psi &=&
  \frac{S}{2}\left\{ \frac{\cos 2\psi}{\cos\Theta}\left[ 1 +
  \epsilon\Lambda_2\sin\Theta
  \right] - \Lambda \right\}+....,
  \label{eq_psi}
\end{eqnarray}
where we define, as in \cite{Lebedev},
\begin{eqnarray}
  S &=& \frac{14\pi}{3\sqrt{3}}\frac{\dot\gamma\eta_0
  r_0^3}{\kappa}\epsilon^{-2}, \\
  \Lambda &=& \frac{23\lambda+32}{240}\sqrt{\frac{30}{\pi}}\,\epsilon,
  \\
  T &=&
  \frac{7}{72}\sqrt{\frac{\pi}{10}}\,\frac{23\lambda+32}{\kappa}\eta_0
  r_0^3\,\epsilon^{-1}
  \label{eq:lebparams}
\end{eqnarray}
and $\Lambda_1 =
  \frac{1}{28}\sqrt{\frac{10}{\pi}}\,\frac{49\lambda+136}{23\lambda+32}$, $\Lambda_2 = 10/7\sqrt{10/\pi}(\lambda-2)/(23\lambda+32)$. The
first term on the right hand side of Eq.~(\ref{eq_alpha})
corresponds to the leading order theory presented in
\cite{Misbah06}. where $....$ stands for other terms of the series.
The first term on the right hand side of Eq.~(\ref{eq_alpha})
corresponds to the leading order theory presented in
\cite{Misbah06}.  The first and second term correspond to the
situation treated in \cite{Lebedev} where only the higher order
contribution in the membrane bending force is
included\cite{Lebedev,Gompper}. Taking the corresponding
hydrodynamical response to the same order into account (as done
here) induce significant changes. A new term, for example, is the
third term on the right hand side of Eq.~(\ref{eq_alpha})
(proportional to $\Lambda_1$). This term is at least of the same
order as  $\cos 3\Theta$. Indeed, the term proportional to $\Lambda
_1 S $ is of the order of $\Lambda _1C_a/\epsilon $.
%Even if we
%consider $\epsilon$ not too small (the situation is worst
%otherwise!), we have a term proportional to $C_a$. If one has in
%mind a physical situation, then it is known that most of
%experimental observations operate at $C_a$ significantly larger than
%one\cite{Mader06,Kantsler06}, and that therefore the neglected terms
%are higher than those retained.
If one has in mind a formal spirit (or a mathematical spirit, in
that $C_a$ is taken of order unity), then $\epsilon$ should  be
regarded as small. In that case the neglected terms are of order
$1/\epsilon$, and are much higher than the retained term in the
Helfrich energy, namely $\cos 3\Theta$ (which is of order one). As a
natural consequence of this, the so-called similarity equations (put
forward in\cite{Lebedev}, in that the evolution equations contain
only 2 independent parameters, $S$ and $\Lambda$; while $T$ can be
absorbed in a redefinition of time) does not hold. Indeed,  we have
three parameters, which are $C_a$, $\lambda$ and $\Delta$, the
excess area (or equivalently $\epsilon$).

 We present now the main outcome following from the study of
Eq.(\ref{eq_ffij}). We first analyse the $tt$ regime. Figure 1
presents the orientation angle as a function of $\lambda$ and
compare the results with previous studies. Instead of a square root
singularity found for the leading order theory (and in the
Keller-Skalak regime\cite{Keller1982}), the angle crosses zero
quasi-linearly. A point which is worth of mention is that the $tt$
angle becomes negative before the solution ceases to exist
(signature of the $tb$ regime). Before the solution ceases to exist
the $vb$ mode takes place, as discussed below.

\begin{figure}
\begin{center}
\vskip-0.3cm
\includegraphics[width=4.5cm,angle=-90]{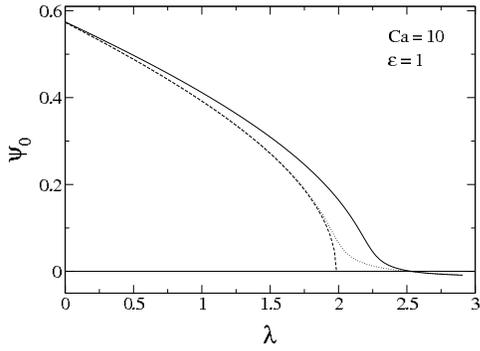}
\caption{The angle $\psi_0$ in the $tt$ regime. Dashed line: the
leading order theory\cite{Misbah06}, dotted line: the theory of
Ref.\cite{Lebedev}, the full line: the present theory.}
\end{center}
\end{figure}

%\begin{figure}
%\begin{center} \vskip-0.3cm
%\includegraphics[height=.8\linewidth,angle=-90]{Fig2.eps}
%\caption{The $tt$-$tb$ boundary.  Dashed line: the leading order
%theory\cite{Misbah06}, dotted line: the KS theory, the full line:
%the present theory.}
%\end{center}
%\end{figure}

%Figure 2 provides the tumbling boundary  in the plane of the reduced
%volume $\nu=[\Delta/(4\pi )+1]^{-3/2}$ and $\lambda$. We find that
%tumbling threshold originally calculated in \cite{Misbah06} is
%significantly improved by the higher order terms.

In \cite{Misbah06} it was predicted that in the tumbling regime a
$vb$ mode should take place. This was found to occur
 as an oscillator (like in a conservative system), since the frequency of
 oscillation about the fixed point $\psi=0$ was found to be purely imaginary.
By including higher order terms the frequency acquires a non zero
real part\cite{Lebedev}, and the $vb$ mode becomes a limit cycle (in
that all initial conditions in its domain of existence tend towards
a closed trajectory in phase space, $(\psi,\alpha)$). As expected
from the original theory\cite{Misbah06} the $vb$ mode still occurs
in the vicinity of the tumbling threshold. This happens provided
that the shape dynamics evolve with time (breathing of the shape).
$C_a$ is a direct measure for the comparison between the shape
evolution
 time scale and the shearing time. The original theory\cite{Misbah06}
 corresponds formally to $C_a=\infty$, as
can be seen from Eq.(\ref{eq_alpha}) and the definition of $T$ and
$S$. Including  higher order terms leads to the appearance of $C_a$
in the equation. In what follows, the definition of $\tau=\eta_{o}
\dot\gamma r_0^3/\bar\kappa$ is adopted.
%that enter the higher order calculation.

\begin{figure}
\begin{center}
\vskip-0.3cm
\includegraphics[height=.8\linewidth,angle=-90]{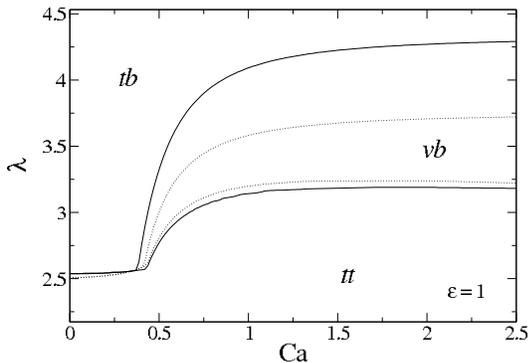}
\caption{The dotted line: theory in Ref.\cite{Lebedev}, the full
line: the present theory. The same order of discrepancy is found
with Ref.\cite{Gompper}.}
\end{center}
\end{figure}

In Fig.2 we report on the phase diagram and compare it to previous
theories\cite{Misbah06,Lebedev}. For small $C_a$ we find a direct
(saddle-node) bifurcation from $tt$ to $tb$, in agreement with
Ref.\cite{Biben03}. At $C_a\rightarrow \infty$ we recover the
results of Ref.\cite{Misbah06} (in that the $vb$ mode coexists with
$tb$ and whether one prevails over the other depends on initial
conditions; this is  not shown on the figure 2). At intermediate
values of $C_a$ we find a belt of $vb$ preceding the $tb$
bifurcation, in qualitative agreement with \cite{Gompper,Lebedev}.
The higher order calculation provided here shows significant
differences with\cite{Lebedev,Gompper}, as shown Fig.2. The results
presented in \cite{Gompper,Lebedev} may be viewed as
semi-qualitative given the disregard of other terms of the same
order. Actually, a simplistic phenomenological model  captures the
main
essential qualitative features of Fig.2\cite{thomas}. %For that purpose we use the
%Keller-Skalak theory\cite{Keller82} and allows for a
%phenomenological equation for the deformability of the ellipsoid.
%The equations take the form:

%THOMAS PART FIGURE  Show the phase diagram and the frequency

Figure 3 shows a snapshot of the $vb$ mode.
%, whereas Fig.4a shows
%the behavior of the long and short axes as a function of time.
Note that a pure swinging (a terminology usually used for
oscillation of rigid objects, and adopted in \cite{Gompper}) would
be impossible within the Stokes limit, since this is forbidden by
the symmetry  of the Stokes equation upon time reversal. The
breathing is a necessary condition for the present mode. In the
upper half plane (i.e. when $\psi>0$) the shape in the $vb$ regime
(dotted line in the Figure) is different from the one in the lower
plane (full line, rounded shape). This asymmetry makes this dynamics
possible owing to the fact that the two shapes (i.e. for $\psi>0$
and $\psi<0$) can not be deduced from each other by a simple mirror
symmetry (albeit this is not easy by necked eyes at first look of
the figure) with respect to the horizontal axis.

\begin{figure}
\begin{center}
\vskip-0.3cm
\includegraphics[width=.5\linewidth,angle=-90]{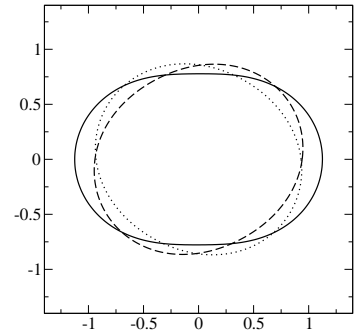}
\caption{A snapshot of the $vb$ mode. Solid line: $\psi=0$. Dotted
line: $\psi\simeq -0.5$. Dashed line $\psi\simeq 0.5$}
\end{center}
\end{figure}

The basic understanding of the $vb$ mode is as follows. First we
recall that  a  shear flow is a sum of a straining part along
$\pm\pi/4$ (which elongates the vesicle for $\psi>0$ and compresses
it for $\psi<0$; see Figure 3) and a rotational part, tending to
make a clockwise $tb$. Due to the membrane fluidity the torque
associated with the shear is partially transferred to $tt$ of the
membrane, so that (due to torque balance) the equilibrium angle for
$tt$ is $0<\psi_0<\pi/4$. Further, an elongated vesicle tumbles more
easily than a compressed one\cite{Biben03}.  Suppose we are in the
$tt$ regime ($\psi_0>0)$, but in the vicinity of $tb$, so
$\psi_0\simeq 0$ . For small $C_a$ the vesicle's response is fast as
compared to shear, so that its shape is adiabatically slaved to
shear (a quasi shape-preserving dynamics): a direct bifurcation from
$tt$ to $tb$ occurs\cite{Biben03}. When $C_a\simeq 1$,  the shape
does not anymore follow adiabatically the shear. When tumbling
starts to occur $\psi$ becomes slightly negative. There the flow
compresses the vesicle. Due to this, the applied torque is less
efficient. The vesicle feels, so to speak, that its actual
elongation corresponds to the $tt$ regime and not to $tb$. The
vesicle returns back to its $tt$ position, where $\psi>0$, and it
feels now an elongation (which manifests itself on a time scale of
the order of $1/\dot\gamma$). Due to  elongation in this position,
tumbling becomes again favorable, and the vesicle returns to
$\psi<0$, and so on. We may say that the vesicle {\it hesitates} or
{\it vacillates} between $tb$ and $tt$. The compromise is the $vb$
mode.
%This motion is possible only due to non-instantaneous
%breathing.

Recently a link between the different modes and rheology has been
presented\cite{Danker07}. It is thus natural to ask how would higher
order terms modify the reported picture.
 Eq.~(\ref{eq_ffij}) constitutes  a basis  for the derivation
of the constitutive law, as in \cite{Danker07}. A full study of the
derivation of the constitutive law will be presented in a future
work. Here we focus on the effective viscosity as a function of
$\lambda$.
%The method of derivation of the effective viscosity follows that
%presented in\cite{Danker07}.
In the $tb$ and $vb$ regimes we make an
average of the effective viscosity over a period of oscillation. The
results are reported on Fig.4

\begin{figure}
\begin{center}
\vskip1.3cm
\includegraphics[height=.5\linewidth,angle=-90]{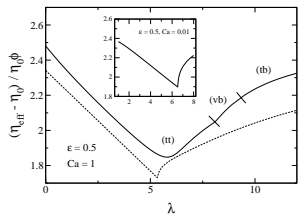}
\caption{ The dashed line in the outer graph corresponds to the
leading order theory where a cusp singularity is observed for any
$C_a$. $\phi$ is the label of the vertical axis is the suspension
volume fraction.}
\end{center}
\end{figure}

\begin{figure}
\begin{center}
\vskip1.3cm
\includegraphics[height=.5\linewidth,angle=-90]{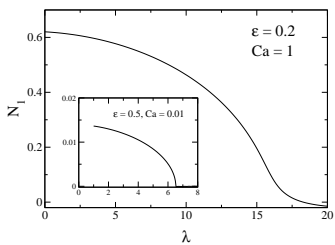}
\caption{The first normal stress difference as a function of
$\lambda$ for two values of $C_a$.}
\end{center}
\end{figure}

We see that at small enough $C_a$, the cusp singularity (inset of
Fig.4)\cite{Danker07} persists as in the leading order theory, while
at larger $C_a$ the cusp is smeared out by the fact that the
transition towards the $vb$ mode  does not show a singularity as
does a saddle-node bifurcation. The same reasoning holds for the
behavior of $N_1$ (Fig.5), the normal stress difference. For low
$C_a$ we recover the calculation in Ref.\cite{Danker07} (that $N_1$
exhibits a square root singularity, while at larger $C_a$ the
behavior is smoother.

We have checked that for high enough $C_a \simeq 100$ ( a quite
accessible value in the experiments\cite{Mader06}) the full
evolution equation produces a coexistence of the $vb$ and $tb$
solution, as predicted by the leading order theory (which is
expected to be recovered formally at high enough $C_a$). We have
checked that depending on initial conditions one mode prevails over
the other (as described in\cite{Misbah06}).

 Finally, a systematic solution of Eq.(\ref{eq_ffij}) reveals new
intriguing  fixed points of the dynamics that do not correspond to
those reported on so far, and they constitute presently an important
line of inquiry. We hope to report along these lines in the future.

C.V. and G.D. gratefully acknowledges support by the European
Commission Marie Curie Research Training Network
MRTN-CT-2004-503661.
 C.M. and T.P. are grateful to CNES (Centre National d'Etudes Spatiales)  and CNRS (ACI ``math\'ematiques
de la cellule et du myocarde'') for a  financial support.

%\newpage
%\noindent Fig. 1: The time dependent orientation angle (upper
%panel), the time dependent viscosity (middle panel) and the first
%normal stress difference (lower panel) in the tumbling regime.

% \noindent
% FIG. 2:
% The reduced average effective viscosity as a function of $h$ for
%the various three regimes: Tank-treading, Tumbling and
%Vacillating-Breathing (VB). The inset shows a cusp singularity at
%the tumbling bifurcation point.

%\noindent FIG. 3: The first normal stress difference $N_1$ as a
%function of $h$ for the various three regimes: Tank-treading,
%Tumbling and Vacillating-Breathing (VB).

\end{document}